\documentclass[aps,twocolumn,prb,superscriptaddress,floatfix,showpacs]{revtex4}
\usepackage{amsmath} 
\usepackage{graphicx} 
\usepackage{subfigure}
\usepackage{color} 
\usepackage{epstopdf}

\def\lsim{\mathrel{\mathpalette\gl@align<}}
\def\gsim{\mathrel{\mathpalette\gl@align>}} \def\gl@align#1#2{\lower.6ex\vbox
{\baselineskip\z@skip\lineskip\z@
\ialign{$\m@th#1\hfil##\hfil$\crcr#2\crcr\sim\crcr}}}
 
\makeatother

\newcommand\ba{\begin{eqnarray}} \newcommand\ea{\end{eqnarray}}
\newcommand\be{\begin{equation}} \newcommand\ee{\end{equation}}
\newcommand\bi{\bibitem}

\begin{document}

\title{Adiabatic dynamics of quasiperiodic transverse Ising model}
\author{Revathy B. S.}
\author{ Uma Divakaran} 
\affiliation{Department of Physics, Indian Institute of
Technology Palakkad, Palakkad, 678557, India}

\date{\today}

\begin{abstract}

We study the non-equilibrium dynamics due to slowly taking a quasiperiodic Hamiltonian
across its quantum critical point. The special quasiperiodic Hamiltonian that we study here
has two different types of critical lines belonging to two different universality classes,
one of them being the well known quantum Ising universality class. In this paper, we verify the
Kibble Zurek scaling which predicts a power law scaling of the density of defects 
generated as a function of the rate of variation of the Hamiltonian. The exponent of this power
law is related to the equilibrium critical exponents associated with
the critical point crossed. We show that the power-law behavior is indeed
obeyed when the two types of critical lines are crossed, with the exponents
that are correctly predicted by Kibble Zurek scaling.
 
\end{abstract}

\pacs{}

\maketitle


\section{Introduction}

Equilibrium phase diagram of systems undergoing quantum phase transitions
(QPT) have been well studied \cite{sachdev99,chakrabarti96,suzuki13}.
Recently, studies on non-equilibrium dynamics of a system which is swept linearly with a speed
$1/\tau$ across a quantum critical point has picked up pace
mainly due to the following two reasons:
(i) the surprising connection between
the non-equilibrium exponent and equilibrium exponents
\cite{polkovnikovRev,dziarmaga10,dutta15},and (ii) the
advancement in the experiments related to optical lattices to simulate quantum
Hamiltonians with very high degree of control and accuracy\cite{bloch08}.  A system 
at zero temperature which is prepared in
its ground state initially, when driven across a quantum critical point by
a time dependent Hamiltonian, gets
necessarily excited.  This is because the relaxation time at the critical point
diverges, implying that the system takes infinite time to respond to the
external variation. Such a situation results to the system not being able to follow the
instantaneous ground state, and hence gets excited.  The density of 
defects $n$ (or excitations)
thus generated is related to the speed of linear variation ($1/\tau$) and
the equilibrium critical exponents by the famous Kibble Zurek (KZ) scaling
$n \sim \tau^{-\frac{\nu d}{\nu z+1}}$, where $d$ is the
dimensionality of the system, $\nu$ and $z$ are the correlation length and time
exponents, respectively, associated with the quantum critical 
point \cite{polkovnikovRev,dziarmaga10,dutta15}. 
The KZ scaling has been verified
in various models and has been accepted as a universal scaling of the density
of defects generated as a result of linear variation of a parameter
of the Hamiltonian \cite{mukherjee07}. 
This scaling can also be generalized to a non-linear variation
of a parameter \cite{mondal09}.

Quasiperiodic lattices have also gained attention recently
as it can have both extended as well as localized states even in one
dimensions \cite{aa,harper}. Till the discovery of quasiperiodic lattices, it was believed
that a disordered Hamiltonian can not have extended states in one or two dimensions,
but can have only localized states. 
It is only in three dimensions or higher that a disordered
system can have extended states as argued by Anderson \cite{anderson58}. 
On the other hand, it is shown that quasiperiodic lattices, which is 
intermediate between periodic and
disordered systems
can still have extended eigenstates \cite{modungo09,deissler11}. The presence of both types of
eigenstates lead to an upsurge in the studies related to quasiperiodic system.
Quasiperiodic Hamiltonians can also be realized in experiments using
lasers of incommensurate lengths \cite{qp_lattice}. Single particle localization in quasiperiodic
lattices has already been observed experimentally \cite{qp_single}.
One such model which has gained a lot of attention is a one-dimensional 
Aubry Andre model which is essentially
an $XX$ spin chain in presence of a quasiperiodic transverse field \cite{aa}.  The beauty
of this model is that it undergoes a phase transition from a phase which
consists of all extended eigenstates to a phase with all localized eigenstates
at a critical value of the transverse field. Moreover, this critical point can
be obtained analytically due to its self dual nature when transformed to
momentum space.\cite{aa,harper}. Theoretical studies on non-equilibrium dynamics
of Aubry Andre model due to sudden and slow variation of a parameter of the Hamiltonian
is studied in Ref. \onlinecite{roosz14}.

Recently, Chandran and Laumann studied a variant of Aubry Andre model, which is Ising
model in presence of quasiperiodic transverse field (QPTIM) \cite{anushya17}. The phase diagram of
this model is richer and involves possibility of having a mobility edge, i.e.,
the system can have both extended and localized eigen states at a particular
value of Hamiltonian parameters and demonstrate the
existence of dynamically stable long range orders which are otherwise 
not present in equilibrium. They showed the existence of localization protected
excited states without disorder. This system also exhibits a new
critical line, the dynamical critical properties of which 
are intermediate between clean Ising critical point and
infinite randomness transition point that arises in disordered model.
The non-equilibrium dynamics generated due to 
sudden quenches of QPTIM is already discussed in Ref. \onlinecite{divakaran18}. 
We now complete the study of QPTIM by looking at
slow variation of a parameter of the Hamiltonian and verify KZ scaling.

This paper is divided as follows: After the basic introduction to the field of KZ scaling
and quasiperiodic Hamiltonian in Section I,
we present a brief description of the Hamiltonian along with the 
proposed phase diagram in Section II. The non-equilibrium dynamics due to adiabatic evolution
of the QPTIM is studied in Section III. We conclude the paper with 
discussions in Section IV.

\section{The model and Phase Diagram}

\begin{figure}[h]
\begin{center} 
\includegraphics[width=8.5cm]{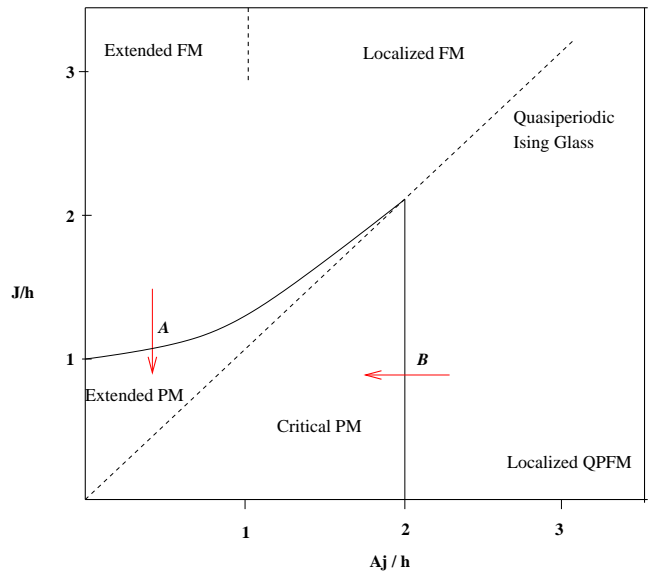} 
\end{center}
\vskip -.5cm \caption{Phase diagram of QPTIM consisting of FM, PM and QPFM
ground states. Depending upon the strength of the quasiperiodic modulation, the
low energy excitations can exhibit localized, extended or critically
delocalized behavior, also shown in the figure.  For more details, see Ref.
\onlinecite{anushya17}.} \label{fig_phasediagram} 
\end{figure}

The Hamiltonian of QPTIM \cite{anushya17} is given by 
\begin{eqnarray} H&=&-\frac{1}{2}\sum_j
J_i\sigma_i^x \sigma_{i+1}^x + h \sigma_i^z,\nonumber\\
J_{i}&=&J+A_J\cos(Q(i+1/2)) \label{eq_ham1} 
\end{eqnarray} 
with $\sigma_i^\alpha$ being the Pauli matrices at site $i$, and $\alpha$
corresponding to $x,y,$ or $z$. 
We choose $Q$ to be the golden ratio $Q=2\pi(\sqrt{5}+1)/2$, so that
the interaction term has a periodicity which
is incommensurate with the lattice.
The zero temperature phase diagram of this
model as obtained by Chandran $et.al$ in $J/h-A_J/h$ plane is presented in
Fig.\ref{fig_phasediagram}. It consists of three phases, namely, 
paramagnet (PM), ferromagnet (FM) and quasiperiodically alternating ferromagnet (QPFM).  
The excited states of the model can be either localized, extended or critical
depending upon the values of $J$  and $A_J$.  The critically delocalized phase
consists of states having multifractal scaling behavior. These properties
of the eigenstates are obtained analytically wherever symmetry permits
and numerically otherwise, the details of which we shall briefly discuss later.
The dashed diagonal line in Fig. \ref{fig_phasediagram} separates two different PM phases based
on their low lying excitations-extended or critical.
The thick line originating from $J/h=1$ corresponds to a phase transition belonging to
the Ising universality class (marked as $A$ in Fig. 1) with $\nu=1$ and $z=1$. 
On the other hand, the second phase boundary separating critical PM and localized 
QPFM belongs to a different universality class (marked B in Fig. 1) 
with the same correlation length exponent as quantum Ising
critical point $i.e.,$ $\nu=1$, but with the dynamical exponent $z$ equal to
$2$.  Therefore, if the adiabatic dynamics involve crossing of the critical
line "A", the defect density will follow
$\tau^{-0.5}$. On the other hand, if the critical line "B"
is crossed, the defect density would give a new exponent with $n \sim
\tau^{-0.33}$. In this work, we verify Kibble Zurek scaling 
while crossing both these different types of critical lines.

For the sake of completeness, we now present a brief discussion on
how the phase diagram of this Hamiltonian is obtained in Ref. \onlinecite{anushya17}.
Owing to the complexity of the Hamiltonian, the form of the eigenstates and corresponding
eigen energies is almost impossible to calculate analytically for all parameter values.
Therefore, the properties of the phase diagram has been obtained analytically
only under certain limits which possess some symmetry, other parts of the phase diagram being
obtained numerically. 
Such special limits include
$A_J=0$, $J=0$ and $J \to \infty$. We consider the limit
of $J \to \infty$ first. It is easy to observe that the ground state of the model 
in this limit consists of all spins pointing either along
$+x$ or $-x$ direction. One can then re-write the Hamiltonian in terms
of domain wall dynamics resembling AA model which helps to extend the information
about AA model to this model.
Such a correspondence allows us to conclude that all the states for $A_J<h$ are extended,
whereas they are localized for $A_J>h$ similar to AA model. 
This particular limit is shown as the 
dashed vertical line in Fig. \ref{fig_phasediagram}. 
Lets consider the next point $A_J=0$ and $J/h=1$
which is the well studied critical point of Transverse Ising model (TIM). It is
well known that the transverse Ising Hamiltonian
has gapless extended excitations at all energies. As seen in Fig. \ref{fig_phasediagram},
there is a parabolic phase boundary originating from TIM critical point and 
extending upto $A_J/h=2$.
 It has been argued using Harris-Luck criterion and also numerically verified that
the weak quasiperiodic modulation is irrelevant at small $A_J$. Therefore, 
this parabolic phase boundary belongs to
the same universality class as that of $A_J=0$, i.e., that of transverse field Ising model.
One expects atleast the low lying excitations along this phase boundary to
be extended.  
On the other hand, at $J =0$ and $A_J/h=2$, there exist a triality similar 
to the AA duality found in Aubry Andre model. Using this triality, it can be shown
that there is a phase transition from critically delocalized to localized states
at all energies at $A_J/h=2$.
All states are localized for $A_J \gg J,h$ \cite{anushya17}.  
It is to be noted that the localization properties are generally claimed to
be $Q$ and energy dependent, which can be cross-checked through numerics. For
more details, please refer to Ref. \onlinecite{anushya17}.

\section{Adiabatic dynamics}

Let us now discuss  the non-equilibrium dynamics generated as a result of linear time
evolution of the transverse field. 
It is to be noted that the phase diagram discussed till now and given 
in Ref. \onlinecite{anushya17} is in
$J-A_J$ plane where the quasiperiodicity is in $A_J$ 
with no quasiperiodicity in $h$.  
In this paper, for the ease of numerical calculations, we have modified the Hamiltonian
bringing the quasiperiodicity in the transverse field, setting $J$ constant and
$A_J=0$. 
The Hamiltonian that we numerically simulate has the following
form
\begin{eqnarray} H&=&-\frac{1}{2}\sum_i
J\sigma_i^x \sigma_{i+1}^x + h_i \sigma_i^z,\nonumber\\
h_{i}&=&h+A_h\cos(Q(i+1/2)) \label{eq_ham2} 
\end{eqnarray} 
which is exactly like Eq. \ref{eq_ham1} with quasiperiodicity in
$J$ shifted to $h$.
It is to be noted that due to the Ising duality in the model, such 
a change will result into 
the paramagnetic and ferromagnetic phases getting swapped leaving the 
dynamical nature of bulk single particle excitations unaltered.  
Quasiperiodicity in $h$ along with $J$ is also discussed in Ref. \onlinecite{crowley18}
by the same authors as in Ref. \onlinecite{anushya17}.
The phase diagram is then in $h/J - A_h/J$ plane
with paramagnet replaced by ferromagnet and QPFM replaced by quasiperiodic 
paramagnet. The critical lines "A" and "B" will still be present 
separating the corresponding similar phases. We shall continue to 
call these critical lines as "A" and "B", though they are now in a different
plane.

We shall first check the non-equilibrium exponent when the quantum Ising 
critical line (marked as A in Fig. \ref{fig_phasediagram}) 
is crossed where the defect density is expected to decay as $\tau^{-1/2}$.
To start with, the system is in its ground state at $t=0$.  
A parameter of the Hamiltonian is varied linearly such that the critical point
is crossed during the evolution.
Close to the
critical point when the relaxation time is larger than the time scale in which
the Hamiltonian is varied, the system is no longer able to follow its ground
state, and gets excited. Below we describe two different quenching protocols
for crossing the above described critical lines. As mentioned before, 
these protocols are presented
in a different plane, i.e., in $h-A_h$ plane as opposed to Fig. \ref{fig_phasediagram}, 
but the new phase diagram will have similar features due to the Ising duality.
To cross the quantum Ising critical line, we divide the evolution
into two steps as follows: (i)start deep in the paramagnet phase 
with $h=5$ and $A_h= 0.1$ and reduce $h$ to zero linearly
as $t/\tau$ which results to crossing of Ising like critical line.
(ii)Next, we reduce $A_h$ to zero with the same rate. 
The final Hamiltonian
after both the steps is simply Ising Hamiltonian with zero transverse field .
Had the evolution been a complete adiabatic evolution, the final state arrived would
be all spins parallel to each other along the Ising direction, 
and any deviation from this would
correspond to defects, which in this case are domain walls. It is easy to see that 
the density of domain walls can be calculated using the following equation \cite{dziarmaga05,caneva07}
\begin{eqnarray} 
n=\frac{1}{L}
\sum_{i}^{L-1}\langle \psi_f|\frac{1}{2}(1-\sigma_i^x \sigma_{i+1}^x)|\psi_f\rangle.  
\label{eq_n}
\end{eqnarray} 
where $|\psi_f \rangle$ is the final evolved state.
The visual realization and ability to write a mathematical
expression for the defects is the main reason for quenching the 
system to Ising Hamiltonian.
As expected, the numerically obtained density of defects after solving Schr\"odinger 
equation follows $n \sim \tau^{-1/2}$ behavior
characteristic of quantum Ising critical point. This is also shown in Fig. 2.
There are two features of this figure. The better agreement with the expected power-law 
for larger system sizes and the faster decay of defect density for  larger $\tau$ values.
This can be attributed to the finite size effects.
The gap is non-zero for a finite system even at critical point. Therefore, it is always 
possible to find some $\tau$ value beyond which the evolution is perfectly adiabatic.
In summary, we need infinite system to get perfect $\tau^{-1/2}$ for all $\tau$ values.

\begin{figure}[h]
\begin{center} 
\includegraphics[width=6.5cm,angle=-90]{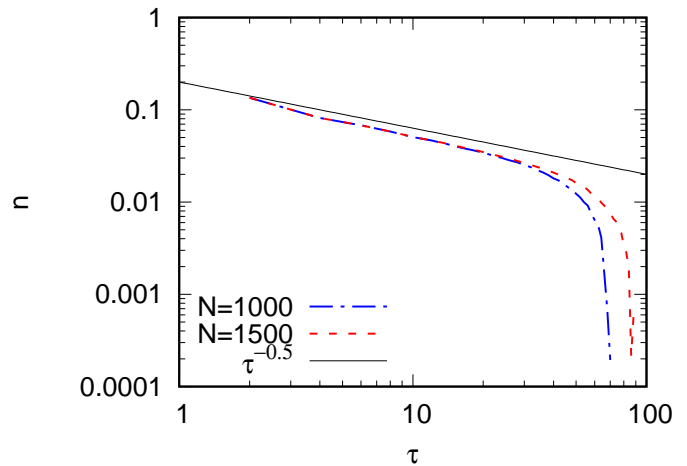} 
\end{center}
\vskip -.5cm 
\caption{Variation of defect density with $\tau$ when the 
critical line "A" is crossed. KZ scaling corresponding to quantum Ising critical
exponents predict $n \sim \tau^{-0.5}$ behavior. As seen above, the defect density
approaches $\tau^{-0.5}$ for larger system sizes.
} \label{fig_ising} 
\end{figure}

\begin{figure}[h]
\begin{center} 
\includegraphics[width=6.5cm,angle=-90]{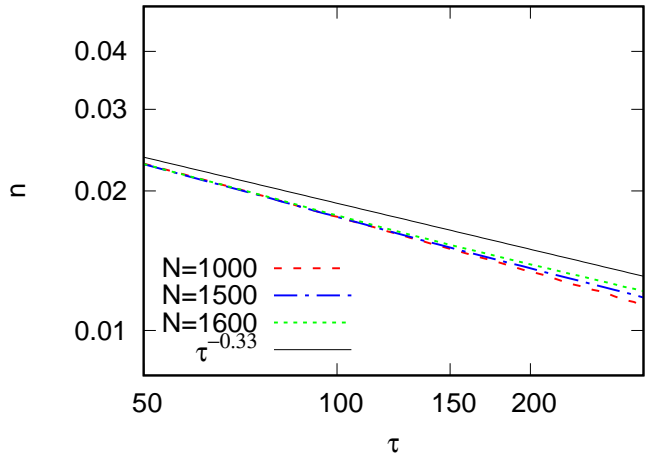} 
\end{center}
\vskip -.5cm 
\caption{Variation of defect density with $\tau$ when the 
critical line "B" is crossed. The defect density $n$ clearly decays with a power
law with the exponent close to $1/3$. The agreement with the exponent increases
as the system size increases. }
 \label{fig_nonising} 
\end{figure}

Let us now look into the exponent when the "B" critical line is crossed.
This is a new type of critical line with a new exponent.
For this, the time evolution of the Hamiltonian is divided into two steps:
(i) Setting $h$ to $0.1$, we reduce $A_h$ as $t/\tau$ 
from $3$ to zero crossing the critical
point.  (ii) In the next step, we reduce the value of $h$ from 0.1
to zero so that at the final time
of the evolution, the Hamiltonian is once again simply Ising Hamiltonian
resulting to same expression for $n$ as given in Eq \ref{eq_n}.  
This protocol ensures that only one critical point belonging to line "B" is crossed.
The critical exponents of the corresponding critical point gives 
a defect density given by $n \sim \tau^{-1/3}$, which is also supported by numerics 
as shown in Fig. \ref{fig_nonising}. In both the figures, one can clearly
see power law behavior, the exponent of which approaches the theoretical 
value as the system size increases.

\section{Conclusion}

We have studied the non-equilibrium dynamics after taking the system through 
two different critical points belonging to two different universality classes.
We find that the conventional Kibble Zurek scaling is obeyed in both the cases.
This work confirms the equilibrium critical exponents obtained in Ref. \onlinecite{anushya17}
for this new model along with the KZ scaling. The numerical calculations clearly show
an improvement in the obtained power law scaling as the system size is increased.
A thorough study of phase diagram with non-zero $A_h$ and $A_J$ has been done
recently in Ref. \onlinecite{crowley18}. It would be interesting to explore
this rich phase diagram in connection with non-equilibrium dynamics also.

\begin{acknowledgments} UD acknowledges DST-INSPIRE Fellowship by
Government of India for financial support.  \end{acknowledgments}


\end{document}